\documentclass[twocolumn, showpacs, preprintnumbers,amsmath,amssymb,prl]{revtex4}

\usepackage{graphicx}
\usepackage{amsmath}

\begin{document}

\title{Mapping Itinerant Electrons around Kondo Impurities}

\author{H.~Pr\"user$^1$}
\author{M.~Wenderoth$^1$}
\email{wenderoth@ph4.physik.uni-goettingen.de}
\author{A.~Weismann$^2$}
\author{R.~G.~Ulbrich$^1$}

\affiliation{$^1$ IV. Physikalisches Institut, Georg-August-Universit\"at G\"ottingen,
Friedrich-Hund-Platz 1, 37077 G\"ottingen, Germany}
\affiliation{$^2$ Institut f\"ur Experimentelle und Angewandte Physik, Christian-Albrechts-Universit\"at zu Kiel, 24098 Kiel, Germany}

\date{\today}

\begin{abstract}
We investigate single Fe and Co atoms buried below a Cu(100) surface using low temperature scanning tunneling spectroscopy. By mapping the local density of states of the itinerant electrons at the surface, the Kondo resonance near the Fermi energy is analyzed. Probing bulk impurities in this well-defined scattering geometry allows separating the physics of the Kondo system and the measuring process. The line shape of the Kondo signature shows an oscillatory behavior as a function of depth of the impurity as well as a function of lateral distance. The oscillation period along the different directions reveals that the spectral function of the itinerant electrons is anisotropic.
\end{abstract}


\maketitle
For a long time it has been well known that a localized spin degree of freedom---a magnetic impurity---in a nonmagnetic host metal significantly alters the scattering behavior of the conduction band electrons of the host as compared to nonmagnetic impurities. This results in a variety of thermodynamic anomalies, which are summarized by the term Kondo effect \cite{Hewson1993}. The most prominent macroscopic hallmark is the resistance minimum at low temperatures observed for metals with magnetic impurities. From a microscopic point of view the impurity interacts with the surrounding electron gas---the itinerant electrons. Below a characteristic temperature, the Kondo temperature $T_\mathrm{k}$, a narrow many-body resonance named Kondo or Abrikosov-Suhl resonance builds up in the spectral function of the impurity at the chemical potential and the impurity spin is effectively screened. Electric transport is dominated by electrons near the Fermi energy for low temperature. Hence the Kondo resonance leads to a strong scattering of electrons at the impurity and an increase of resistivity with decreasing temperature.

Microscopic properties of single impurity Kondo systems on the atomic scale regained interest by recent scanning tunneling microscopy (STM) experiments \cite{Madhavan1998, Manoharan2000, Schneider2002, Knorr2002, Quaas2004, Zhao2005} on single magnetic ad-atoms and molecules deposited on noble metal surfaces (for a review see \cite{Ternes2009}). In these works the Kondo effect manifests itself as a sharp signature in the STM differential conductance around zero bias. Studies which investigate the dependence of the Kondo signature on the lateral distance \cite{Madhavan1998, Knorr2002} showed that the signal rapidly vanishes and the line shape nearly remains constant. This is in contrast to theory \cite{Ujsaghy2000,Schiller2000,Plihal2001} which predicts a long range visibility and oscillatory behavior of the spectral function of the itinerant electrons - the local density of states (LDOS). Models for tunneling through ad-atom systems have to treat the tip, its coupling to the localized impurity state, its coupling to the itinerant bulk electrons as well as surface states, and all interferences between alternative transport paths \cite{Schiller2000,Plihal2001}. 

During recent years STM was refined to investigate nano structures \cite{Schmid1996, Kurnosikov2008, Kurnosikov2009, Brovko2010, Kurnosikov2011} or even single impurity atoms \cite{Weismann2009, Sprodowski2010, Avotina2010, Lounis2011} which are buried below a metal surface. The interference pattern on the surface caused by the sub surface scattering centers are determined by the band structure of the underlying host material \cite{Weismann2009}. This induces strong directional Friedel oscillations with wave vectors and amplitudes depending on the crystal direction \cite{Lounis2011}. Recently, it was shown that for magnetic impurities buried below the surface a long range signature of the Kondo effect is observable \cite{Prueser2011}.

In this letter we present a detailed analysis of differential conductance as a function of lateral distance for Fe and Co atoms situated at different depths below a Cu(100) surface. We show that the distance dependence of the line shape of the Kondo signature is directly connected to the band structure of the host crystal. In contrast to previous works using ad-atom systems which had studied only local properties, this work focuses on single Kondo impurities in the bulk on length scales, which hitherto were not accessible. The long range Kondo signature allows us to explore the full microscopic properties around magnetic impurities. Additionally, probing bulk impurities with STM allows separating the physics of the correlated many-body system and the measuring process.

The experiments were performed using a home-built low-temperature scanning tunneling microscope (STM) operating at 6K at pressures below $5\cdot10^{-11}$ mbar. The Cu(100) single crystal substrate is cleaned by several cycles of argon bombardment and electron beam heating. A copper alloy with a small amount ($\approx 0.02\%$) of impurities is prepared through simultaneous deposition of copper and the impurity material from two electron beam evaporators.

\begin{figure}[ht]
\begin{center}
\includegraphics[angle=0,width=7.5cm]{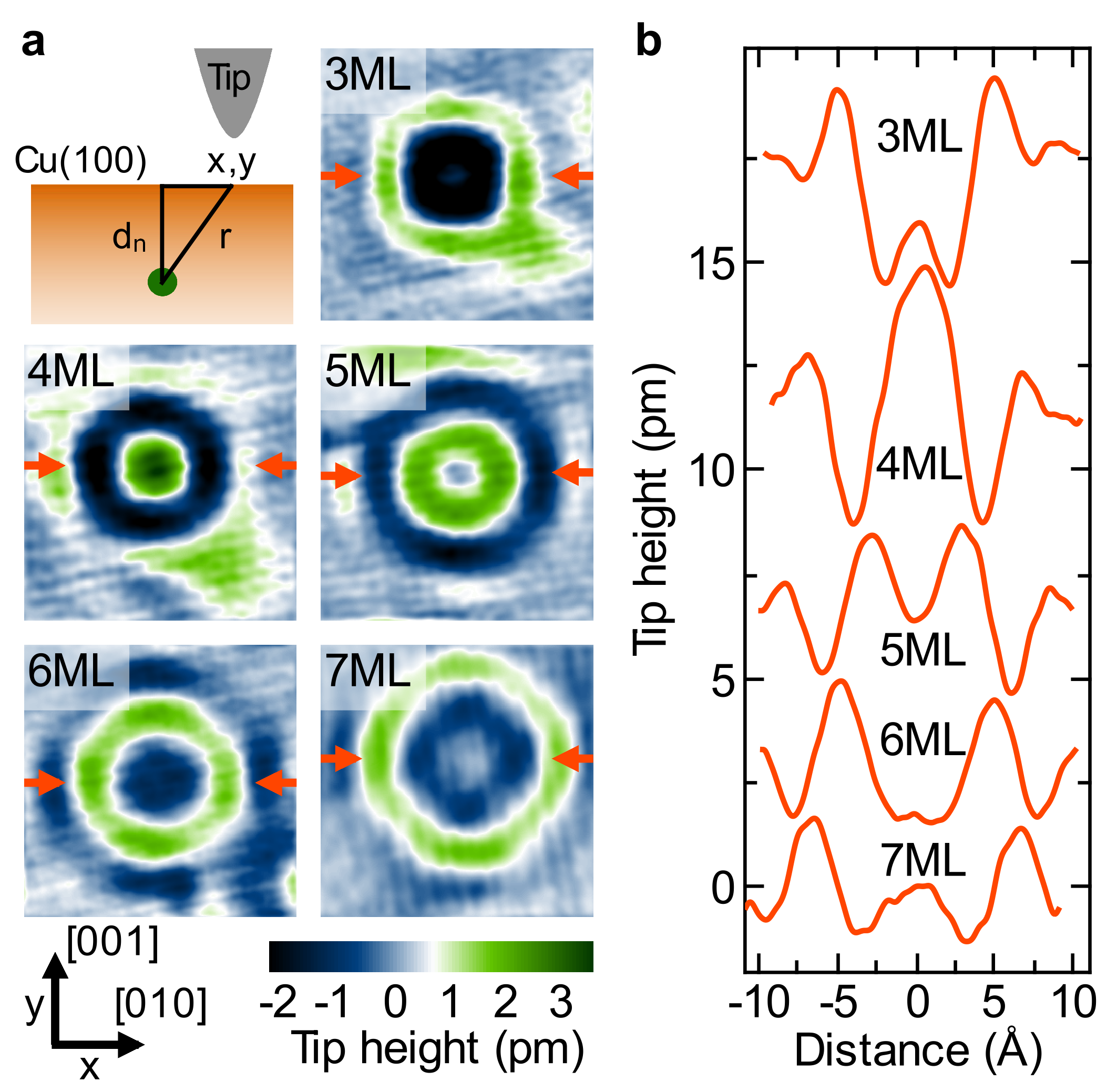}
\caption{(a) Scheme of the experiment: the green circle marks the impurity position below the surface and constant current topographies (2nm x 2nm, V = 10mV, I = 2nA) of single Co atoms buried below a Cu(100) surface. (b) Corresponding cross sections (shifted vertically for clarity) of each layer along the x = [010] direction, marked by the arrows in (a).}\label{fig1}
\end{center}
\end{figure}

The embedded atoms are identified by their short period standing-wave pattern which has four fold symmetry \mbox{[Fig.\,\ref{fig1}a]}. The modulation in the topographic images has a wavelength of $5.5 \pm 0.4$\AA$ $ and the amplitude is of the order of a few pm \mbox{[see Fig.\,\ref{fig1}b]}. Since the surface is atomically flat and the applied bias voltage is small, the modulation amplitude in the topographic images can be directly connected to a change in the local density of states LDOS. Assuming a work function of 4.3eV, a change of 1pm in tip-sample distance results in a change in the LDOS of 2\%. Consequently, the LDOS modulations that cause the observed modulation amplitudes are in the range of  6\% (7ML) - 13\% (3ML). The depth $d_\mathrm{n}$ expressed in monolayer (ML) of each impurity below the surface is determined as follows: the observed standing wave patterns are ordered by the lateral size. The size of the standing wave pattern increases monotonically with depth of the impurity. As a verification of the obtained impurity position we use atomically resolved topographies. Since the Cu(100) has an fcc crystal structure and impurity atoms are located on lattice sites, the impurity contrast follows a certain pattern with respect to the host lattice. We label the surface layer as 1st monolayer. The standing wave pattern of an odd layer impurity is centered on a corrugation maximum of the surface. If the impurity is positioned in an even layer the center is located between the corrugation maxima.

To investigate the embedded impurities and their influence on the LDOS at the surface we use scanning tunneling spectroscopy (STS) data by recording an $I\!\left(V\right)$ curve at every scanning point with interrupted feedback loop. Further data processing includes averaging and numerical differentiation. This provides a complete map of the differential conductance $\mathrm{d}I/\mathrm{d}V\!\left(x,y,V\right)$ as a function of lateral tip position $\left(x,y\right)$ on the surface and applied bias voltage $V$. In order to perform a quantitative analysis of the experimental data, drift effects which occur during acquisition of spectroscopic data are removed by comparing the simultaneously measured topography with a reference topography obtained in the normal constant current mode.

\begin{figure}[ht]
\begin{center}
\includegraphics[angle=0,width=7.5cm]{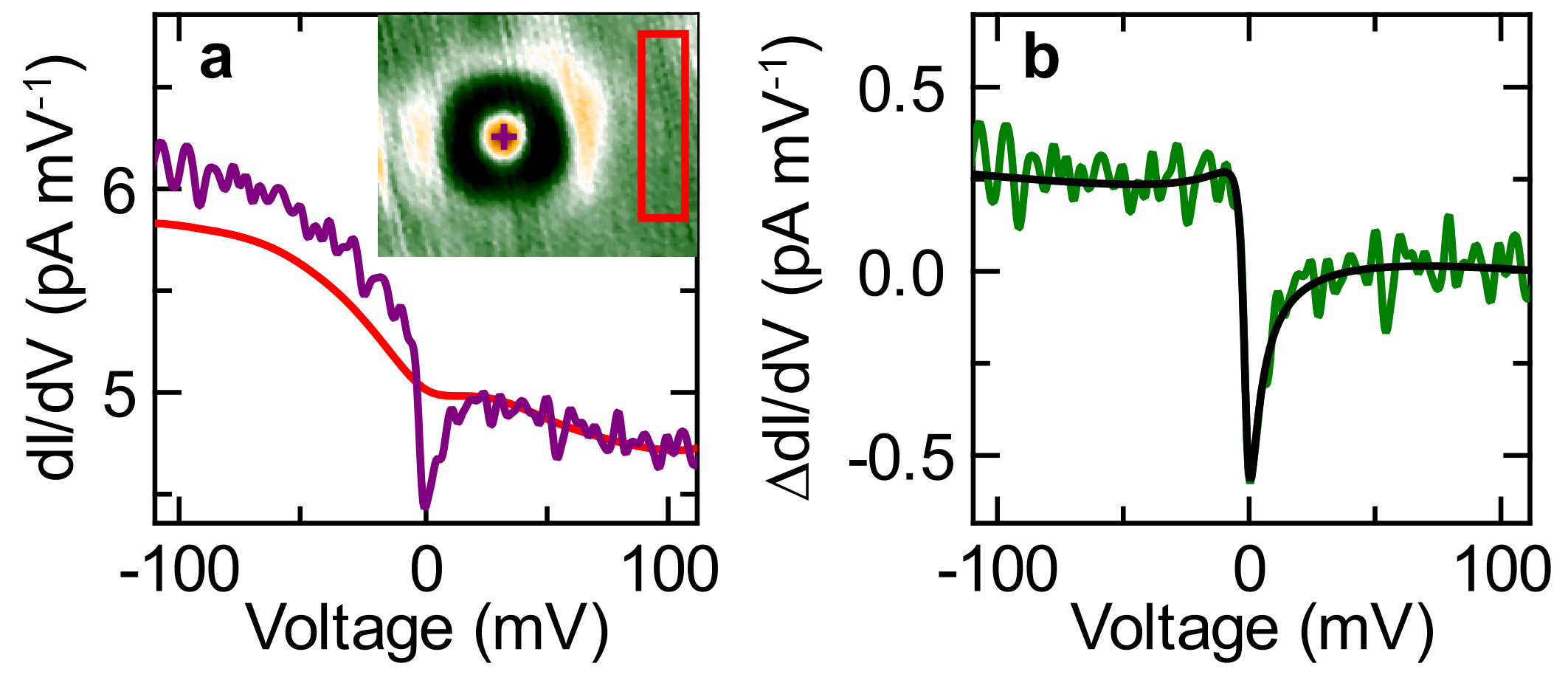}
\caption{(a) Spectroscopy of a 3ML Fe impurity, the corresponding topography (2.5nm x 2.0nm, V = -312mV, I = 1.8nA) is shown in the inset. A pair of $\mathrm{d}I/\mathrm{d}V$ spectra taken with the STM tip held over the center (purple curve) of the interference pattern, evaluated at the purple cross and over the nearby bare Cu surface (red curve) evaluated above the undisturbed region of the image (red rectangle). (b) Difference of the differential conductance $\Delta \mathrm{d}I/\mathrm{d}V = \mathrm{d}I/\mathrm{d}V - \mathrm{d}I_0/\mathrm{d}V$ (green curve) above the impurity ($\mathrm{d}I/\mathrm{d}V$) and above the free surface ($\mathrm{d}I_0/\mathrm{d}V$). The solid black curve shows a fit to the $\Delta \mathrm{d}I/\mathrm{d}V$ data described in the text.}\label{fig2}
\end{center}
\end{figure}

Typical differential conductance $\mathrm{d}I/\mathrm{d}V$ spectra measured directly at the center of the interference pattern reveal a strong signature around zero bias voltage which is attributed to the Kondo effect and cannot be seen in spectra of the bare Cu surface \mbox{[see Fig.\,\ref{fig2}a]}. To extract the LDOS change and to remove artifacts which originate from the bare Cu(100) surface and the tip, we normalize the obtained differential conductance around an impurity by subtracting the spectra of the free surface $\mathrm{d}I_0/\mathrm{d}V$ far away from the impurity. The difference is proportional to the change in the LDOS at the lateral tip position \cite{Wahl2008}. The normalized  $\Delta \mathrm{d}I/\mathrm{d}V$ spectrum \mbox{[Fig.\,\ref{fig2}b]} only shows a dip around zero bias, all other signatures coming from the free surface and tip are removed.

Although the Kondo effect has been subject of a wealth of experimental and theoretical investigations, microscopic real space descriptions are rare. The most relevant property is the spectral function of the localized orbital, which has been directly measured by high-resolution photo emission electron spectroscopy and inverse photo emission \cite{Reinert2005}. In an STM experiment the current is determined by tunneling into the conduction electron density of states. Therefore, the STM is probing the spectral function of the itinerant electrons and not the localized ones. A first theoretical description of the Kondo effect in the spectral function of the itinerant electrons at different distances from the impurity was given by  \cite{Ujsaghy2000,Schiller2000,Plihal2001}, more recent results can be found in \cite{Affleck2008, Buesser2010}. They all showed that the Kondo resonance leads to a scattering resonance in the LDOS resulting in an enhanced scattering amplitude and an energy-dependent phase shift at the Fermi energy \cite{Prueser2011}. Here we show, that the Kondo signature shows an oscillatory behavior as a function of depth of the impurity as well as function of lateral distance. Furthermore we analyze the influence of the electronic band structure of the host material on the line shape of the Kondo signature. As described in \cite{Ujsaghy2000} the LDOS at different positions depends on the propagation of the conduction band electrons between tip and impurity (a property of the host metal) and the scattering behavior of the impurity, which is strongly influenced by the Kondo resonance. We replace the Lorentzian shape of the Kondo resonance by a phenomenological form found by Frota et al. \cite{Frota1986, Frota1992}. Their formula gives a better description of STM experiments \cite{Fu2007, Prueser2011} and is justified by theoretical many-body numerical renormalization group (NRG) calculations \cite{Zitko2011}. The final expression for Frota's phenomenological line shape has the following functional form:
\begin{align}
     \frac{\Delta \mathrm{d}I}{\mathrm{d}V} = {a}\cdot\mathrm{Im} \left[i e^{i\phi}\sqrt{\frac{i \Gamma}{eV - e_{\mathrm{k}}+i\Gamma}\,} \right] + {b}\cdot V + {c}
\label{eqn_fit}
\end{align}
Here $e_{\mathrm{k}}$ is the position of the Kondo resonance. The resonance width $\Gamma$ is proportional to the Kondo temperature $T_{\mathrm{k}}$, whereas the exact prefactor is still an open issue. We use $\Gamma = 1.43\,k_{\mathrm{B}}\,T_{\mathrm{k}}$, which results from NRG calculations \cite{Zitko2009}. A linear voltage slope is added in Eq.~(\ref{eqn_fit}) to account for additional and approximately energy independent background scattering processes. Measurements taken on different impurity depth and spatial positions yield a resonance width $\Gamma\mathrm{(Fe)} = 4 \pm 2$meV and $e_{\mathrm{k}}\mathrm{(Fe)} = 0 \pm 3$meV for Fe impurities and $\Gamma\mathrm{(Co)} = 81 \pm 19$meV and $e_{\mathrm{k}}\mathrm{(Co)} = -65 \pm 25$meV for Co atoms. Our observation that for Co the Kondo resonance is situated well below the Fermi energy is in good agreement with recent calculations \cite{Surer2012}.

\begin{figure}[ht]
\begin{center}
\includegraphics[angle=0,width=7.5cm]{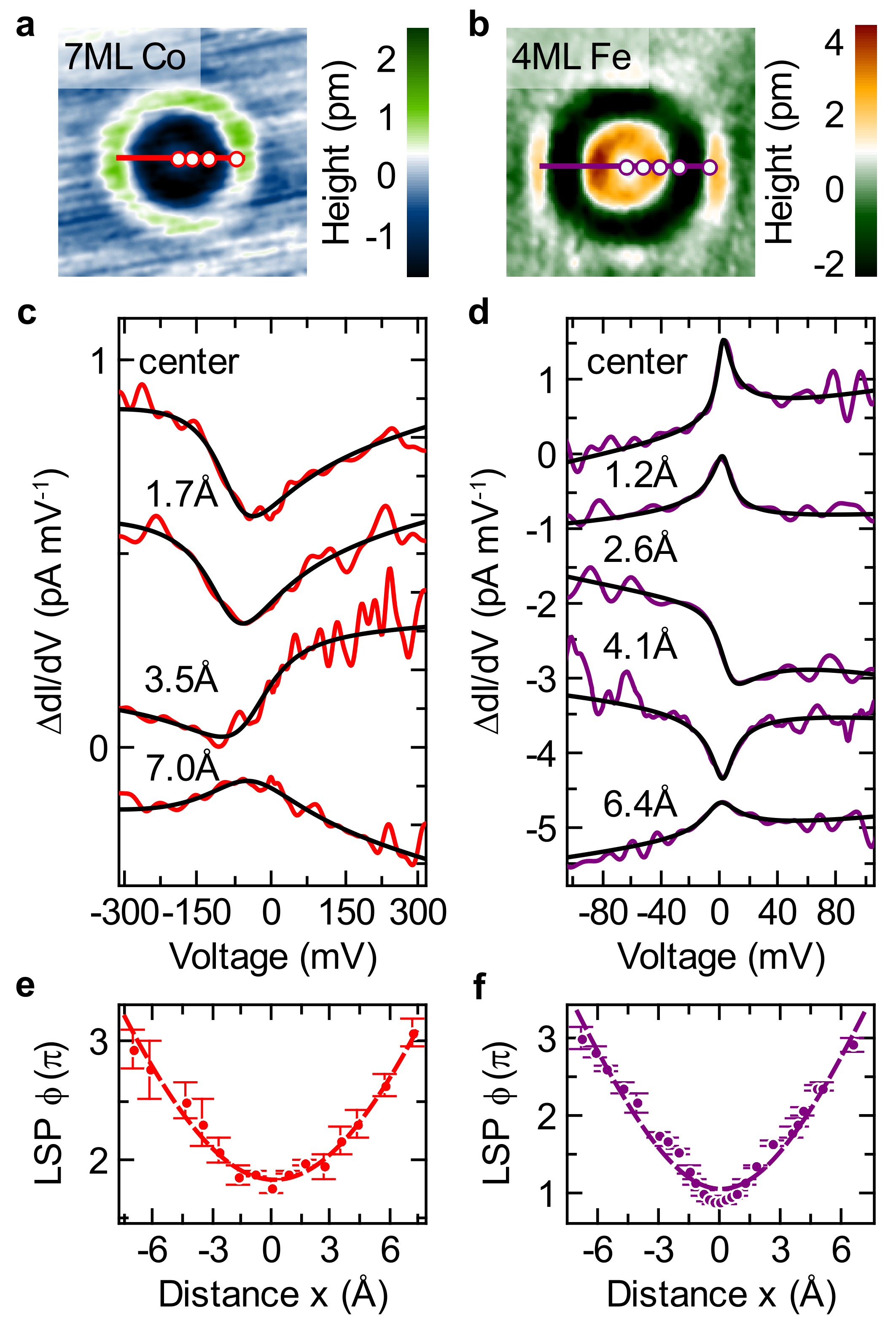}
\caption{Line-shape variation as a function of the lateral tip position. (a) STM constant current topography of a 7th layer Co impurity (3nm x 3nm, V = -300mV, I = 2nA) and (b) a 4th monolayer Fe impurity (2nm x 2nm, V = -150mV, I = 2nA), in which the tip position for the different spectra are depicted. Single spectra, displaced vertically for clarity, for different lateral distances, measured from the center along the [010] direction (c) for the 7ML Co impurity and (d) for the 4ML Fe impurity. The solid black curve shows fits to the $\Delta \mathrm{d}I/\mathrm{d}V$ described in the text. Spatial dependence of the line-shape parameter (LSP) for 7ML Co (e) and for 4ML Fe (f). Error bars are obtained by the confidence interval of the non linear fit. The dashed lines show a model presented in the text.}\label{fig3}
\end{center}
\end{figure}

In the following we will show that for a magnetic atom the line shape of the $\mathrm{d}I/\mathrm{d}V$ spectra strongly depends on the depth of the impurity as well on the lateral distance \cite{Note1}. The line shape is determined by $\phi$ in Eq.~(\ref{eqn_fit}). The line-shape parameter (LSP) results for even multiples of $\pi$ ($\phi = 0, 2\pi,...$) in a dip and for odd multiples of $\pi$ ($\phi = \pi, 3\pi,...$) in a peak.

First, we analyze the dependence on the lateral position x (the [010] direction, see sketch in \mbox{Fig.\,\ref{fig1}a}) while keeping the depth $d_\mathrm{n}$ of the impurity fixed. To illustrate this, the topographies of a 7th Co and 4th layer Fe impurity are shown, respectively in \mbox{Fig.\,\ref{fig3}a} and \mbox{Fig.\,\ref{fig3}b}. A series of $\Delta \mathrm{d}I/\mathrm{d}V$ spectra taken with the STM tip held at various lateral spacing from the center of the interference pattern is shown in \mbox{Fig.\,\ref{fig3}c} and \mbox{Fig.\,\ref{fig3}d}. At the center $\left(x,y = 0\right)$ directly above a 4th layer impurity \mbox{[Fig.\,\ref{fig3}d]} a peak is observed, which corresponds to a value of $\phi = 0.83\pi$. Moving away from the center the line shape becomes more asymmetric, turns into a negative peak and back into a symmetric peak at a distance of $x = 6.4$\AA. The line shape parameter increases continuously with the lateral position \mbox{[Fig.\,\ref{fig3}f]}. The same behavior is observed for a Co atom \mbox{[see Fig.\,\ref{fig3}e] }.

The line shape parameter should depend linearly \cite{Ujsaghy2000} on the distance $r$ to the impurity with a slope given by 2 times the Fermi wave vector $k_F$, i.e., $\phi\left(r\right) = 2\,k_\mathrm{F} \, r$. The distance $r = \sqrt{d_{\mathrm{n}}^2 + x^2}$ in our experiment is determined directly from the lateral distance $x$ and the depth $d_{\mathrm{n}}$ of the impurity below the surface. We use the following relation $\phi = 2\,k_{\mathrm{F}}\,\sqrt{d_{\mathrm{n}}^2 + x^2}+\phi_0$ to fit the line-shape parameter. The offset $\phi_0$ accounts for the line shape directly at the impurity position. For a 4th layer Fe impurity ($d_4 = 5.4$\AA) the best fit is obtained with parameters $\phi_0 = -0.2 \pm 0.4\pi$ and $k_{\mathrm{F}} = 0.98 \pm 0.08$\AA${^{-1}}$. For a 7ML Co atom ($d_7 = 10.8$\AA) the best parameters are $\phi_0 = 0.1 \pm 0.8\pi$ and $k_{\mathrm{F}} = 0.85 \pm 0.15$\AA${^{-1}}$.

All theoretical models so far describing the line-shape dependence as a function of distance treat a homogeneous and isotropic electron gas model with a direction independent $k_{\mathrm{F}}$. However, band structure calculations for Cu yield values of $k_{\mathrm{F}} = 1.3$\AA$^{-1}$ in [110] and $k_{\mathrm{F}} = 1.45$\AA$^{-1}$ in [100] direction \cite{Segall1962}. The difference between our experimental and the theoretical value is not only observed for the two data sets shown here, but for all our experimental findings \cite{Note2}. The discrepancy can be basically explained by two additional ingredients: the anisotropic band structure of copper, i.e., the focusing effect and the imaging process of the STM. The noncollinear direction of $k_{\mathrm{F}}$ and the group velocity results in an effective reduction of the observed oscillation wave vector \cite{Lounis2011}. In addition, not all wave functions contribute equally to the tunnel current. Especially states with higher $k_{||}$ components decay faster into the vacuum. This implies that for larger lateral distances of the tip states with different $\left|k_{\mathrm{F}}\right|$ contribute to the tunnel current.

\begin{figure}[ht]
\begin{center}
\includegraphics[angle=0,width=5.5cm]{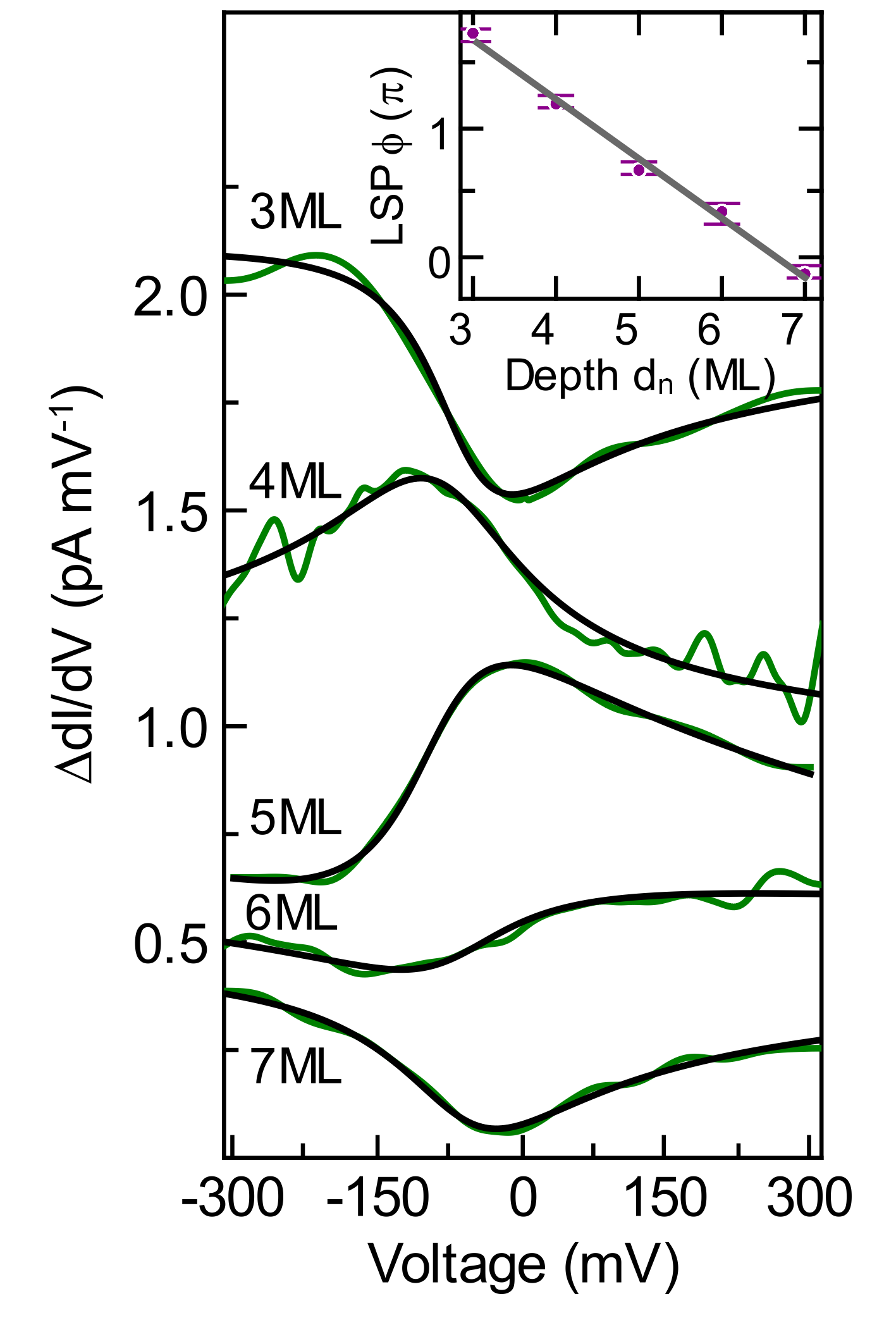}
\caption{Experimentally determined line-shape parameter (LSP) for Co atoms as function of the impurity position $d_{\mathrm{n}}$ below the surface (the corresponding topographies are shown in Fig.\,\ref{fig1}). Measured $\Delta \mathrm{d}I/\mathrm{d}V$ (green) and calculated spectra (black) with a tip position fixed at the center of the interference pattern. Inset: The obtained LSP (purple dots) decreases linearly. Error bars are obtained by the confidence interval of the non linear fit.}\label{fig4}
\end{center}
\end{figure}

A way to analyze the line-shape dependence in a simpler geometry is to look at spectra with different impurity depth $d_{\mathrm{n}}$ while the lateral tip position is kept constant at the center $\left(x,y = 0\right)$ of the interference patterns \mbox{[Fig.\,\ref{fig4}]}. In this geometry the contribution of tunneling paths is the same for all observed spectra. The spectrum for a 3ML Co atom reveals an asymmetric dip. For deeper impurity positions the line shape becomes more asymmetric returning into an asymmetric dip for the 7ML, seen before for the 3ML, indicating an oscillation length of nearly 4ML. The line-shape parameter \mbox{[inset of Fig.\,\ref{fig4}]} apparently decreases linearly with increasing depth. A linear fit reveals a slope of $-0.45 \pm 0.02 \pi$/ML. This negative slope is due to aliasing between the periodicity of the lattice and the wave length of the oscillations. The line shape parameter is 2$\pi$ periodic and hence the observations are also equivalent to a slope of $1.55 \pm 0.02\pi$/ML, which results in an Fermi wave vector of $k_{\mathrm{F}} = 1.35 \pm 0.05$\AA${^{-1}}$. This is in good agreement with the expected value of $k_{\mathrm{F}} = 1.45$\AA$^{-1}$ for copper in [100] direction \cite{Segall1962}.

Finally, we would like to discuss our results in comparison to previous results on ad-atom systems. In the description of ad-atom spectra the line shape of the Kondo resonance is discussed in terms of interfering channels considering tunneling into the localized orbital and the conduction band states, respectively. Thus, the line shape is a feature of the measurement process and is associated with the relative strength of the hybridization between the tip, sample and impurity. In contrast, the investigation of bulk impurities allows a clear spatial separation between the Kondo impurity and STM process. Within our interpretation, the STM is a noninvasive tool that exclusively probes the LDOS of the conduction band electrons---the itinerant electrons.

In conclusion, our results show that STM is able to characterize bulk impurities on relatively long distances from the impurity, which opens a new way to Kondo physics in real space. Using STS we have demonstrated that single magnetic Fe and Co impurities embedded below the Cu(100) surface show a complex spatially dependent Kondo signature. The spatial distribution of the Kondo signature is---in addition to the Kondo physics---strongly influenced by the anisotropy of the band structure of the host crystal. This work was supported by the Deutsche Forschungsgemeinschaft through SFB 602 Project A3.

\bibliographystyle{unsrt}

\end{document}